\documentclass[twocolumns]{aa} 
\usepackage{graphicx}
\def\gtrsim{\buildrel > \over {_{\sim}}}
\def\lesssim{\buildrel < \over {_{\sim}}}

\begin{document}
   \title{Gamma Rays from Cosmic Rays in Supernova Remnants}


   \author{C.\ D.\ Dermer
          \inst{1}
          \and
          G.\ Powale\inst{1,2}
\thanks{Currently at the University of California, Berkeley}
	}

   \institute{Space Science Division, Code 7653,
              U.\ S.\ Naval Research Laboratory, 4555 Overlook Ave., SW, Washington DC 20375 USA\\
              \email{charles.dermer@nrl.navy.mil}
         \and
             Broad Run High School, Ashburn, VA 20147 USA\\
             \email{gauripowale@yahoo.com}
             }

   \date{Received 17 September 2012; accepted 18 March 2013}

 
  \abstract
   {Cosmic rays are thought to be accelerated at supernova remnant (SNR) shocks, but obtaining conclusive evidence for this hypothesis is difficult. }  
   {New data from ground-based $\gamma$-ray telescopes and the Large Area Telescope on the Fermi Gamma-ray Space Telescope are used to test this hypothesis. 
A simple model for $\gamma$-ray production efficiency is compared with measured $\gamma$-ray luminosities of SNRs, and the GeV to TeV fluxes ratios of SNRs are examined for correlations with SNR ages.}
   {The supernova explosion is modeled as an expanding spherical shell of material
that sweeps up matter from the surrounding interstellar medium (ISM). The accumulated kinetic energy of the shell, which provides the energy available for nonthermal particle acceleration, changes when 
matter is swept up from the ISM and the SNR shell decelerates. A fraction of this energy is assumed to be converted into the energy of cosmic-ray electrons or protons.  Three different particle radiation processes---nuclear pion-production interactions, nonthermal electron bremsstrahlung, and Compton scattering---are considered.  }
   {The efficiencies for $\gamma$-ray production by these three processes are compared with $\gamma$-ray luminosities { of SNRs. Our results suggest that SNRs become less $\gamma$-ray luminous} {\rm at $\gtrsim 10^4$ yr, and are consistent with the hypothesis that supernova remnants accelerate cosmic rays with an efficiency of $\approx 10$\% for the dissipation of kinetic energy into nonthermal cosmic rays. Weak evidence for an increasing GeV to TeV flux ratio with SNR age is found.}}
   {}


\authorrunning{Dermer \& Powale}
   \maketitle
%

\section{Introduction}

Supernovae have been hypothesized (Ginzburg \& Syravotskii \cite{gs64}, Hayakawa \cite{hay69}) 
to be the sources of the cosmic rays, which consist primarily of relativistic protons, ions and electrons. 
Firm evidence for this hypothesis is lacking because cosmic rays 
are deflected by intervening magnetic fields and do not trace back to their sources. 
A distinctive $\gamma$-ray signature of cosmic rays results from
interactions of cosmic-ray protons and ions with gas and dust in the vicinity of their acceleration sites, 
making $\gamma$ rays 
with a characteristic $\pi^0\rightarrow 2\gamma$ emission feature peaking at 67.5 MeV 
in a photon number representation (Stecker \cite{ste71}). 

Resolution and sensitivity limitations of previous $\gamma$-ray telescopes made it impossible
to precisely localize $\gamma$-ray sources in the Galaxy and to measure the expected $\pi^0$-decay spectral
signature. This has changed with the new generation of $\gamma$-ray telescopes, including
AGILE (Astro-rivelatore Gamma a Immagini LEggero; Gamma-ray Light Imaging
Detector) and the Fermi Large Area Telescope (LAT). Evidence for a cutoff below
several hundreds of MeV from some SNRs has been found (Giuliani et al.\ \cite{giu11}, 
Abdo et al.\  \cite{AbdoW44}, \cite{AbdoIC443}, Ackermann et al. \cite{2013Sci...339..807A}), so
this one-hundred year old problem may soon be solved.

While spectral analyses at photon energies $E_\gamma \lesssim 100$ MeV continue, 
we can also ask whether the measured $\gamma$-ray luminosities of SNRs  are consistent
with expectations of efficiency for cosmic-ray acceleration at SNR shocks. Furthermore, trends in 
the $\gamma$-ray spectra might reveal a progressive
behavior characteristic of cosmic-ray acceleration in these sources.

Here we test the hypothesis that cosmic rays are accelerated at SNRs
by calculating the fraction of supernova kinetic energy that 
must be converted into cosmic-ray particles in order to produce 
the $\gamma$ rays that are observed 
from some of these remnants. We also plot the GeV to TeV energy-flux ratio with age.


\section{Particle acceleration at SNR shocks}

Supernovae (SNe) are found in two main classes, namely thermonuclear Type Ia SNe and core-collapse Type II SNe.\footnote{Other subclasses
of core-collapse SNe subclasses are Type Ib and Ic SNe.}  
The progenitors of Type Ia SNe are Solar mass white dwarfs composed of carbon and oxygen. 
Thermonuclear burning of the material in the white dwarf releases a prodigious amount of energy and creates a shock wave
as the burned material expands. 
For stars more than $\approx 8$ times as massive as the Sun, in contrast, the stellar interior burns until much of it
has been converted to an Fe core, which will collapse when electron-degeneracy pressure is no longer sufficient
to support the core against gravity.  A core bounce sends a shock wave out that 
expels a portion of the stellar envelope (for a general review, see Burrows \cite{bur00}). 
Neutron stars can be formed in Type II SNe, but not in Type Ia SNe. 

In both Types Ia and II SNe, the energy available for nonthermal particle production is derived from the 
directed bulk kinetic energy of the shell. This energy is dissipated in the form of the random kinetic energy of matter swept up
at the SNR shock. The amount of energy that is extracted and transformed into cosmic-ray energy is highly
uncertain, and represents a major open question in SNR modeling. Different prescriptions related
to the number of swept-up particles or the kinetic energy dissipated at the shock front have been considered
(e.g., Sturner et al.\ \cite{stu97}, Reynolds \cite{rey98}, Baring et al.\ \cite{bar99}, Tang et al.\ \cite{tan11}), 
but the exact conversion efficiency depends on poorly understood microphysical processes (e.g., Blasi et al.\ \cite{bla05}).
Here we reconsider this problem using an idealized model of an expanding spherical shell of matter 
and a uniform surrounding medium (cf.\ Truelove \& McKee \cite{tm99}; 
Finke \& Dermer \cite{fd12}).

Let  $E_0 = 10^{51}E_{51}$ erg represent the total kinetic energy of the supernova ejecta 
with mass $M_0 = m M_\odot$, where $M_\odot = 2.0\times 10^{33}$ g is the mass of the Sun.  
For Type Ia and II SNe, $m \cong 1.4$ and $m \cong 10$, respectively, 
with $E_{51}\approx 1$ for both types. The SN is here
assumed to explode in a medium of uniform density $n_0$ and form an outward moving 
spherical shell of material with radius $r = r(t)$ at 
time $t$ after the explosion. As it expands, the shell sweeps 
up material with mass $M_{su}(t) = 4\pi r^3 \rho_0/3$, where the density $\rho_0 = m_p n_0$,
and the ISM is assumed to be composed of hydrogen. 
The initial speed of the shell is therefore $v_o = \sqrt{2E_0/M_0} \cong 10^4 \sqrt{E_{51}/m}$ km s$^{-1}$. 

The Sedov radius $r_{\rm Sed}$ is defined by the condition that the swept-up mass equals the explosion mass, that is,
$4\pi \rho_0 r_{\rm Sed}^3/3 = M_0$, implying $r_{\rm Sed} = (3M_0/4\pi\rho_0 )^{1/3} = 6.6\times 10^{18}(m/n_0)^{1/3}$ cm.
The Sedov time $t_{Sed}$ is the time when the shell transitions from the coasting phase at $r < r_{\rm Sed}$ to 
the Sedov phase at $r> r_{Sed}$, and is given by $t_{Sed} = r_{\rm Sed}/v_0 = 210\, m^{5/6}E_{51}^{-1/2} n_0^{-1/3}$ yr.

For nonrelativistic speeds $v_o\ll c$ and adiabatic shock waves, the kinetic energy 
\begin{equation}
E_{ke} = {1\over 2} [M_0 + M_{su}(t)]v^2
\label{Eke}
\end{equation}
is conserved, so that 
$d E_{ke}(t)/dt = 0$.   
Introducing dimensionless radius $x = r/r_{\rm Sed}$ and time $\tau = t/t_{\rm Sed}$ in units of Sedov radius and Sedov time, 
respectively, and noting that  $v = dr/dt$, Eq.\ (\ref{Eke}) leads to the equation of motion 
\begin{equation}
(1+x^3)^{1/2} dx = d\tau
\label{eqofmotion}
\end{equation}
for the dimensionless shell radius $x$. From this, the asymptotes 
\begin{equation}
x(\tau ) \rightarrow \left\{ \begin{array}{ll}
    \tau\;, & \tau \ll 1 \\
    \left(5\tau /2\right)^{2/5}\;, & \tau \gg 1 
  \end{array}
  \right. \ 
\label{xtau}
\end{equation}
and
\begin{equation}
v(\tau ) \rightarrow \left\{ \begin{array}{ll}
    v_0\;, & \tau \ll 1\;, \\
    v_0\left(5\tau /2\right)^{-3/5}\;, & \tau \gg 1 
  \end{array}
  \right.  \
\label{vtau}
\end{equation}
can be derived.

   \begin{figure}
   \centering
   \includegraphics[width=3.0in]{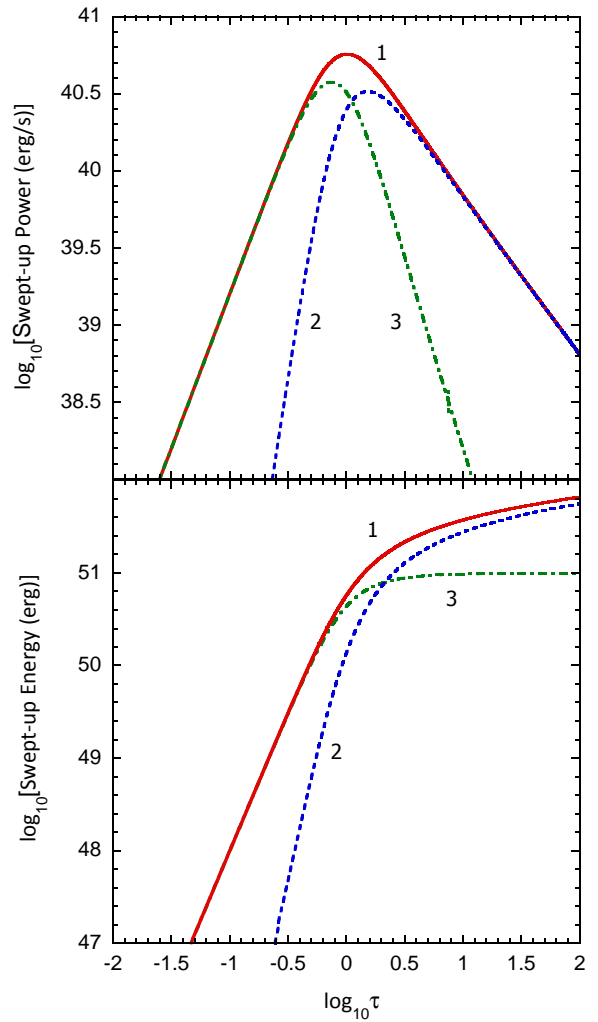}
      \caption{{\it Top:} Red curve labeled ``1" shows the rate at which kinetic energy is swept into the 
SNR shell, and the blue curve labeled  ``2" shows the rate at which kinetic energy 
is lost due to deceleration of the shell.  The net rate of change of kinetic energy in the SNR shell is 
given by the green curve, labeled ``3". {\it Bottom:} Integrated kinetic energy corresponding to the three terms shown
in the upper panel. Curve 3 gives the accumulated swept-up kinetic energy, ${E}_{su}(t)$. 
              }
         \label{powerenergy}
   \end{figure}

If nonthermal particles extract their energy from the kinetic energy of swept-up matter, then the change of kinetic energy of 
swept-up matter with time is given by 
$${d E_{ke}^{su}(t)\over dt} = {d\over dt}\left[ {1\over 2} M_{su}v^2\right ] ={1\over 2}\dot M_{su}v^2 + M_{su}v \dot v\;$$
\begin{equation}
=2\pi r^2 v^3 \rho_0 + {4\pi \over 3}r^3 v\dot v \rho_0\;.
\label{dEkesut}
\end{equation}
The first term on the right-hand-side of Eq.\ (\ref{dEkesut}) is the rate at which kinetic energy is swept into the shocked material, 
and the second term is the rate at which the kinetic energy of the swept-up matter is lost by deceleration of the shell (and is therefore
negative). The sum represents the net rate of change of kinetic energy in the SNR shell. These terms are shown in the top 
panel of Fig.\ \ref{powerenergy}, and the integrated energy associated with each term is shown in the bottom panel of 
Fig.\ \ref{powerenergy}. Here we let $E_{51} = 1$, $m = 1.4$, and $n_0 = 0.1$ cm$^{-3}$. Integrating the swept-up kinetic energy 
leads to a logarithmically diverging total energy that exceeds the initial 
kinetic energy of the explosion. By adding the change of energy associated with deceleration, the summed contribution recovers
the initial kinetic energy of the explosion at late times. 

The process of sweeping up matter from the surrounding medium has 
transformed the directed kinetic energy into internal thermal and nonthermal kinetic energy of swept-up matter. { First-order Fermi acceleration at astrophysical shocks provides a specific mechanism to convert directed kinetic energy into nonthermal particle kinetic energy (e.g., Blandford \& Eichler  \cite{be87}), although here we treat the dissipation mechanism through a hypothetical conversion efficiency. }  Note
how rapidly, $\propto \tau^{-2.5}$, that the total injection rate of kinetic energy decreases in the Sedov phase.


\section{Cosmic-ray interactions in the SNR environment}

Cosmic-ray ions and electrons can interact with target particles to make $\gamma$ rays through secondary nuclear production and 
bremsstrahlung, respectively. Additionally, relativistic electrons will Compton-scatter photons of the 
ambient radiation fields to the $\gamma$-ray regime (see Blumenthal \& Gould \cite{bg70} and Dermer \& Menon \cite{dm09} for reviews of high-energy
astrophysical radiation processes). There are three targets available for particle-particle interactions, namely from (1) the explosion mass,
(2), the swept-up mass, and (3) the ISM gas. The explosion mass density $n_{ex}(t) = M_0/m_pV_{sh}(t)$,  where
the volume of the expanding shell of width $\Delta$ is $V_{sh}(t) = 4\pi[r^3-(r-\Delta )^3]/3 \rightarrow 
4\pi r^2 \Delta$ in the limit $u \equiv \Delta/r \ll 1$. Hence 
\begin{equation}
n_{ex}(t) = {M_0\over 4\pi r^3 u m_p} = {n_0\over 3x^3 u}  \rightarrow {n_0\over 3u}\left\{ \begin{array}{ll}
    \tau^{-3}\;, & \tau \ll 1 \\
    (5\tau/2)^{-6/5}\;, & \tau \gg 1 \;.
  \end{array}
  \right. \ 
\label{xtau}
\end{equation}
The density of the swept-up mass is increased by the compression ratio. For a strong shock,
the shell of swept-up matter has { density $n_{su} \cong 4n_0$.} 

Cosmic rays accelerated at the forward shock and convecting downstream into the shocked fluid would interact with the compressed ISM with density $\approx 4n_0$.
The most efficient radiation production
 occurs when the cosmic rays interact with explosion mass in the coasting phase, and swept-up mass
in the Sedov phase. These two regimes can be bridged with a formula for the maximum target density 
available for interactions, given by
\begin{equation}
n_{max}(t) \cong {n_0\over 3ux^3}\,(1+12ux^3)\;.
\label{nmaxt}
\end{equation}
For cosmic rays to interact with the dense SNR shell material would require rapid diffusion
from the region downstream of the forward shock into the expanding SNR shell.   We consider the case where
the target is the shocked ISM and where the CRs interact with the densest available target.

Secondary nuclear production makes $\gamma$ rays from the decay of neutral pions, as well as through 
emissions from pion-decay electrons and positrons. For cosmic-ray protons with energies $E_p = m_p c^2 \gamma_p$ well above the pion-production threshold energy at $E_p \cong 300$ MeV, the rate at which the proton Lorentz factor $\gamma_p$ changes is given by $-\dot \gamma_{pp} = K_p c\sigma_{pp} n(t)\gamma_p$, where $K_p\approx 0.5$ is the mean inelasticity (fraction of energy lost per collision), and $\sigma_{pp} \approx 30$ mb is the p-p inelastic cross section, so that 
the energy-loss time scale associated with secondary production is  
$t_{pp}  = |\gamma_p/\dot\gamma_{pp}| = [K_p c\sigma_{pp} n(t)]^{-1}\cong 2.2\times 10^{15}/n(t)$ s. The maximum $\gamma$-ray luminosity
that can be made by secondary nuclear production is therefore 
\begin{equation}
L_{pp}(t) \cong { \eta_{pp} \,{E}_{su}(t)\over 3 t_{pp}(t)},
\label{Lppt}
\end{equation}
with $n(t)$ replaced by $n_{max}(t)$, from Eq.\ (\ref{nmaxt}), in $t_{pp}(t)$. The factor $1/3$ represents the mean fraction of 
energy lost in an inelastic nuclear interaction that emerges as $\gamma$ rays, and the term  $\eta_i$ is the fraction of accumulated
swept-up kinetic energy  that is transformed into cosmic-ray protons { or electrons,  where $i = pp, ff,$ and {\it C} stand for nuclear production, 
electron bremsstrahlung, and electron Compton scattering, respectively}.

The bremsstrahlung energy-loss rate for electrons with energy $E_e = m_ec^2 \gamma_e$ interacting with ions of density $n_Z$ and charge
$Z$ in a fully ionized medium is given by $-\dot \gamma_{ff} = k_{ff}\alpha_f c\sigma_{\rm T}[\Sigma n_Z Z(Z+1)]\gamma_e$, 
where $\alpha_f$ and $\sigma_{\rm T}$ are the fine-structure constant and Thomson cross section, respectively. 
The term $k_{ff} = (3/2\pi)(\ln 2\gamma_e - {1\over 3})  \approx 4.3 $, taking $\gamma_e \approx 6000$ for the typical 
electron Lorentz factor $\gamma_e$ that makes GeV $\gamma$ rays through this process. The characteristic bremsstrahlung
energy-loss time scale is therefore $t_{ff}(t) = |\gamma_e/\dot\gamma_{ff}|\cong 8.0\times 10^{14}/n(t)$ s, and the maximum bremsstrahlung luminosity is  
\begin{equation}
L_{ff}(t) \cong { \eta_{ff} \,{E}_{su}(t)\over t_{ff}(t)},
\label{Lfft}
\end{equation}
again replacing $n(t)$ with $n_{max}(t)$ in $t_{ff}(t)$. 

The loss rate of cosmic-ray electrons by Compton scattering is given by 
$-\dot\gamma_{C} = (4/3)c\sigma_{\rm T} U_\gamma \gamma_e^2/m_ec^2$, 
where $U_\gamma$ is the energy density of the { target radiation} field.
The characteristic energy-loss time scale through Compton { scattering in the Thomson limit} is therefore
$t_{C}$(s)$ = |\gamma_e/\dot\gamma_{C}| = 7.7\times 10^{19}/\gamma_e $ $ \approx 7.1\times 10^{13}/\sqrt{E_{\rm GeV}}$, 
where we assume that the dominant radiation field is the cosmic microwave background (CMB) radiation 
with energy density $U_\gamma = U_{CMB} \cong 4.0\times 10^{-13}$ erg cm$^{-3}$. 
In the final expression, $\gamma_e$ is replaced by the value of the electron Lorentz factor that will scatter a typical 
CMB photon to a $\gamma$ ray with energy $E_{GeV}$ GeV. The $\gamma$-ray luminosity from Compton scattering is therefore
\begin{equation}
L_{C}(t) \cong { \eta_{C} \,{E}_{su}(t)\over  t_{C}(t)}.
\label{LCst}
\end{equation}
{ The nonthermal electrons will additionally lose energy through synchrotron losses, with total synchrotron luminosity
larger than $L_C$ by a factor $U_B/U_\gamma$, where $U_B$ is the magnetic-field energy density. The magnetic field is constrained
in SNR spectral modeling, but does not affect the $\gamma$-ray emission model presented here. }


  \begin{figure}
   \centering
   \includegraphics[width=3.0in]{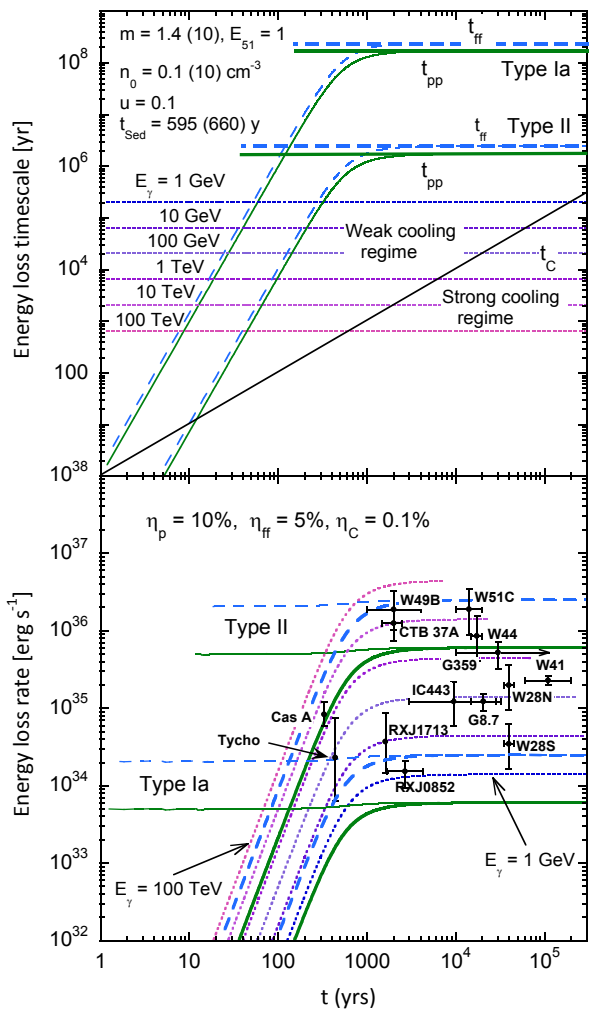}
      \caption{{\it Top:} Energy-loss time scales of cosmic-ray protons through secondary nuclear production ($t_{pp}$; solid curves) and of cosmic-ray electrons through bremsstrahlung ($t_{ff}$; dashed curves) are shown for parameters corresponding to Type Ia and Type II SNe, as labeled (values for Type II SNe are given in parentheses in the legend). Also shown are time scales for Compton energy losses ($t_{\rm C}$; dotted curves) of cosmic-ray electrons that scatter CMB photons to make $\gamma$ rays of energy $E_\gamma$. {\it Bottom:} Theoretical estimates of $\gamma$-ray luminosities of SNRs through secondary nuclear production (solid curves), bremsstrahlung (dashed curves), and Compton scattering (dotted curves). Data points show the measured $> 300$ MeV $\gamma$-ray luminosities for the SNRs listed in Table 1. Thick curves in the two panels are for cosmic-ray interactions with the shocked external medium, and thin curves are for cosmic-ray interactions with the densest available target material.  Model curves are plotted only in the weak-cooling regime.
              }
         \label{time scales}
   \end{figure}

The curves shown in Fig.\ \ref{time scales} give results for the energy-loss time scales in the top panel, and the $\gamma$-ray luminosities of SNRs in the bottom panel.\footnote{We calculate a $\gamma$-ray luminosity for CTB 37A larger, by an order of magnitude, than Castro and Slane (\cite{cs10}).} For the model results, we assume that Type Ia and II SNe involve 1.4 and 10 $M_\odot$ of explosion mass, and that the surrounding medium densities are $0.1$ and 10 cm$^{-3}$, respectively. The total kinetic energy of the explosion is assumed to be $10^{51}$ erg in both cases, and we let $u = 0.1$ for the relative width of the shell to the radius.  The Sedov time scales for these parameters are $595$ yr and $660$ yr for the Type Ia and Type II SNRs, respectively. Also shown are the energy-loss time scales for cosmic-ray electrons that scatter CMB photons to $\gamma$ rays of energy $E_\gamma$. 
The solid line in the upper panel separates the weak and strong cooling regimes; in the strong-cooling regime our treatment does not apply. 

The lower panel of Fig.\ \ref{time scales} gives model results for the $\gamma$-ray luminosities using the same parameters as shown in the upper panel. The results are plotted only in the slow-cooling regimes where the model applies. The swept-up kinetic energy is assumed to be transformed into cosmic-ray protons and electrons with efficiencies of $\eta_{pp} = 10\%, \eta_{ff} = 5$\%, and $\eta_{C} = 0.1$\% for the three processes. The efficiency for converting kinetic energy into electrons is small for the latter process, but these electrons have to have $\gamma_e \cong 10^6 \sqrt{E_{GeV}}$, { large by comparison with bremsstrahlung $\gamma$ rays made by electrons with $\gamma_e \sim 3 E_{GeV}/m_ec^2 \sim 6\times 10^3 E_{GeV}$.  

The injection efficiency should not, however, depend on the $\gamma$-ray emission process.  { In the case of a very soft electron spectrum, the characteristic amount of energy injected into lower-energy, $\gamma_e \sim 10^4$ electrons making bremsstrahlung would be much greater than the energy injected into the higher-energy, $\gamma_e\sim 10^6$ electrons that Compton-scatter soft photons to GeV energies, and this could account for the different efficiencies that are implied by the data for the two processes.} This would, however, conflict with the hard electron spectra required to model SNRs. At 10 GeV, where Solar modulation effects and energy losses on cosmic-ray protons and electrons are small, the ratio of the cosmic-ray electron to proton fluxes is $\approx 1\%$. If SNRs are the sources of the cosmic-ray protons and electrons, then the injection efficiency for { electrons should therefore be much less than for protons, and a bremsstrahlung origin of the $\gamma$-ray emission could be ruled out   (Yuan et al. \cite{ylb12})}. 

Note that the increase in swept-up kinetic energy competes with the decrease in target density so that the $\gamma$-ray luminosity is roughly constant with time for secondary nuclear production and bremsstrahlung when using the target density given by eq.\ (\ref{nmaxt}). The $\gamma$-ray luminosity through nuclear production { and bremsstrahlung} declines at early times when cosmic-ray protons interact only with the swept-up ISM with density $\approx 4n_0$. The target photon density for Compton scattering is constant, so the $\gamma$-ray luminosity through this process tracks the increasing swept-up kinetic energy shown in Fig.\ \ref{powerenergy}, and is also therefore smaller at early times.

   \begin{table}
      \caption[]{$\gamma$-ray SNRs.$^{\mathrm{a}}$}
         \label{GeVTeVSNRs}
     $$ 
         \begin{array}{lllll}
            \hline
            \noalign{\smallskip}
            {\rm SNR}  & {\rm Type} &  {\rm Age~(kyr)} & {\rm d~(kpc)}  & {\rm Refs.}^{\mathrm{b}}  \\
            \noalign{\smallskip}
            \hline
            \noalign{\smallskip}
            {\rm Cas~A} & {\rm II}	& 0.33    &   3.4^{+0.3}_{-0.1}  &   1 \\
           {\rm  Tycho}  &  {\rm Ia}      & 0.44  & 1.5 - 4.0   & 2 \\
	    {\rm RX~J1713.7-3946}& {\rm II(?)}	& 1.6	&  0.9 - 1.7   & 3,4	\\
	    {\rm  RX~J0852.0-4622}  &  {\rm II(?)} & 1.7 - 4.3   &   0.75   & 5,6   \\
	    {\rm CTB~37A}& {\rm II}	& 2\pm 0.5	&  6 - 9.5  &  7,8	\\
	    {\rm W49B}	& {\rm II}	& 1 - 4	&  8 - 11 & 9 \\
           {\rm  IC443}  &  {\rm Ib(?)} & 3 - 30^{\mathrm{c}}   &   \sim 1.5   & 10   \\
	   {\rm  W51C}	& {\rm II}	&  10 - 30&  7\pm1.5 & 11	\\
           {\rm G359.1-0.5}  &  {\rm } 		& > 10   &  7.6   & 12,13  \\
	    {\rm W44}	& {\rm II}	&  15 - 20	& \sim 3    &  14,15 \\
           {\rm  G8.7-0.1}  &  {\rm II}{^\mathrm{e}} 		& 15 - 28  &   4.8 - 6.0   &16,12   \\
	   {\rm  W28}	& {\rm II}		&  35 - 45 &  1.9\pm0.3	& 17,18\\
           {\rm  W41}  &  {\rm II}^{\mathrm{d}} 		& 60 - 200   &   3.9 - 4.5  & 19,20   \\

            \noalign{\smallskip}
            \hline
         \end{array}
     $$ 
\begin{list}{}{}
\item[$^{\mathrm{a}}$] Except for W44, all SNRs are detected at TeV energies.
\item[$^{\mathrm{b}}$] 1. Abdo et al.\ \cite{AbdoCasA}; 
2. Giordano et al.\ \cite{gio12}; 
3. Abdo et al.\ \cite{AbdoRX}; 
4. Fesen et al.\ \cite{fes12}; 
5. Tanaka et al.\ \cite{tan11}; 
6. Aharonian et al.\ \cite{aha07};
7. Brandt  \cite{bra13}; 
8. Tian et al.\ \cite{tia12};
9. Abdo et al.\ \cite{AbdoW49B};
10. Abdo et al.\ \cite{AbdoIC443}; see also Caprioli \cite{cap11};   
11. Abdo et al.\ \cite{AbdoW51C}; 
12. Aharonian et al.\ \cite{aha06};
13. Hui et al.\ \cite{hui11};
14. Abdo et al.\ \cite{AbdoW44}; 
15. Wolszczan et al.\ \cite{wol91};
16. Ajello et al.\ \cite{aje12};
17. Abdo et al.\ \cite{AbdoW28}; 
18. Vel{\'a}zquez et al.\ \cite{vel02}; 
19. Leahy \& Tian \cite{lt08};
20. Tian et al.\ \cite{tia07}

\item[$^{\mathrm{c}}$] Troja et al.\ \cite{tro08} argue from XMM-Newton X-ray observations that the age is $< 10$ kyr.
\item[$^{\mathrm{d}}$] { This shell-type SNR is probably a core-collapse SNR given its proximity to molecular clouds. The TeV radiation might also be powered by a pulsar wind, in which case this analysis would not apply; see Mukherjee et al.\ \cite{mgh09}.
\item[$^{\mathrm{e}}$] { Possibly associated with PSR 1800-21; see Kassim \& Weiler \cite{km90}. TeV emission could also be powered by the pulsar wind.}}
\end{list}
   \end{table}

\section{Comparison with data} 

 We calculated the $\gtrsim 300$ MeV $\gamma$-ray luminosities $L_\gamma$ from the data provided in the references for the SNRs listed in Table 1. The error bars in $L_\gamma$ reflect both the uncertainties in the $\gamma$-ray measurements as well as distance uncertainties. The values of $L_\gamma$ range from $\approx 10^{34}$ erg s$^{-1}$ to $\approx 3\times 10^{36}$ erg s$^{-1}$. The data show an apparent increase in luminosity from young to { intermediate-aged ($\sim 3$ ky)} SNRs, followed by a decline for the { middle-aged} ($\gtrsim 10$ kyr) SNRs, but a larger sample of $L_\gamma$ from SNRs will be needed to confirm this trend. Most of these SNRs are Type II, so that if the parameters used to characterize the Type II SNRs are reasonable, then we find that $\eta\sim 10$\% for nuclear processes { to account for the $\gamma$-ray luminosities of the most luminous SNRs}, generally consistent with expectations of production efficiency if SNRs accelerate the cosmic rays. {  There are, however, sources that deviate significantly from the $\eta= 10$\% curves, for example, W28. Precise determinations of production efficiency requires an examination of SNRs on a case by case basis, because the surrounding medium density and composition, ejected mass and kinetic energy of the SNR, not to mention the limitations of the simple model, will introduce a wide range of luminosities even in the case that $\eta$ is constant.}

Four young, $\lesssim 3000$ yr old SNRs have $\gamma$-ray luminosities $\lesssim 10^{35}$ erg s$^{-1}$. Tycho is a Type IA in a region of low density. If cosmic rays interact only with the shocked material, though not with the SNR shell, then the luminosity of Cas A is in accord with this simple model (see Fig.\ 2). { SNRs RX~J1713.7-3946 and RX~J0852.0-4622}, though they may be Type II SNRs, also take place in low density regions, which will diminish the relative contribution from nuclear processes.

In our simple model, SNRs from Type II SNe formed in dense environments are predicted to be $\approx 2$ orders of magnitude more $\gamma$-ray luminous than Type Ia SNe, and therefore more easily detected. { Given the greater rate of Type II than Type Ia SNe in the Galaxy (e.g., Cappellaro et al. \cite{cap99}), this would explain the larger number of core-collapse SNRs in our sample. The assumed larger ISM densities in the vicinity of core-collapse SNe are expected because Type II and Type Ib/c SNe are typically found in star-forming regions. Indeed, OH maser emissions from CTB 37A suggests that it is interacting with a very dense, $n_0 > 10^3$ cm$^{-3}$, molecular cloud (Hewitt et al.\ \cite{hyw08}; Castro \& Slane \cite{cs10}). The ratio of nuclear production and Sedov time scales is $\sim 8\times 10^4 E_{51}^{1/2}/m^{5/6} n_0^{2/3}\propto n_0^{-2/3}$ for cosmic rays interacting with the swept-up ISM, so Sedov-age SNRs in dense environments would shine more brightly through nuclear interactions than Compton processes  (cf.\ Yuan et al. \cite{ylb12}) than those in low-density surroundings. The SNRs RX~J1713.7-3946 and RX~J0852.0-4622 are, however, found in tenuous, $n_0 <0.1$ cm$^{-3}$ environments  (Slane et al.\ \cite{sla99}, \cite{sla01}), which could happen when core-collapse SNe take place in stellar wind cavities. Their hard spectra favors a Compton-origin for the gamma rays.  }

  \begin{figure}
   \centering
   \includegraphics[width=3.0in]{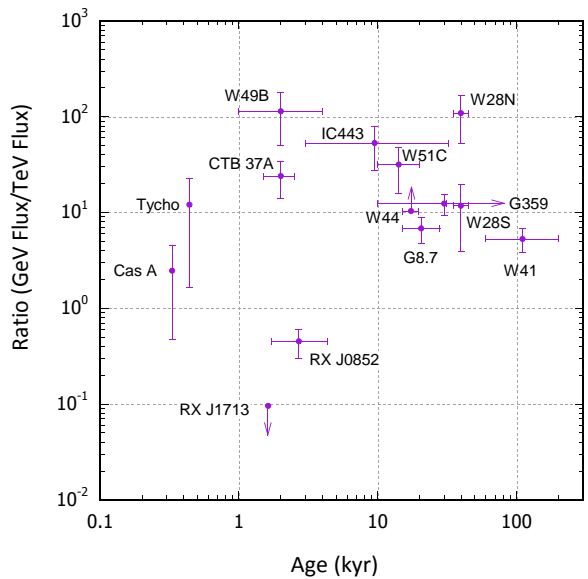}
      \caption{Ratio of 0.3 -- 3 GeV to 0.3 -- 3 TeV fluxes for the SNRs given in Table 1.  
              }
         \label{GeVTeVfluxes}
   \end{figure}

We also plotted the ratio of the 0.3 -- 3 GeV energy flux to the 300 GeV -- 3 TeV energy flux for the SNRs listed in Table 1. The result is shown in Fig.\  \ref{GeVTeVfluxes}. All of the SNRs except for RX~J1713.7-3946 and RX~J0852.0-4622 have GeV to TeV flux ratios of order unity or greater. The middle-aged SNRs are more luminous in GeV $\gamma$ rays than in TeV $\gamma$ rays by factors of $\gtrsim 10$ -- 100. RX~J1713.7-3946 is anomalously low in comparison. There is a weak trend for increasing GeV to TeV flux ratio with age. This might be expected if the dominant processes are Compton scattering and nuclear production, noting from Fig.\ \ref{time scales} that cosmic-ray electrons making TeV photons lose their energies on time scales of $\sim 10^{4}$ yr, leaving the long-lived cosmic-ray protons that would preferentially make GeV radiation. 

Compton scattering as a multi-TeV $\gamma$-ray emission process for middle-aged SNRs might be ruled out by noting how rapidly the high-energy electrons that scatter CMB photons to TeV energies cool (Fig.\ \ref{time scales}). If these electrons are powered by swept-up kinetic energy, energetic electrons in the strong cooling regime would not be replenished due to the rapid decrease in the kinetic-energy injection rate shown by curve 3 in Fig.\ \ref{powerenergy}.   Compton production of 10 -- 100 TeV $\gamma$-rays requires electrons with $\gamma_e \sim 10^7$ -- $10^8$, irrespective of target photon energy. Thus the Compton scattering process can be ruled out by detection of $\gg$ TeV radiation from SNRs much older than the Sedov age, provided that it can be shown that there is no pulsar wind to inject high-energy electrons. Indeed, a rapid decline in the X-ray synchrotron radiation from $\gamma\sim 10^8/\sqrt{(B/10~\mu{\rm G})}$ electrons shortly after the Sedov phase is found by plotting the nonthermal X-ray luminosity as a function of radius (Nakamura et al.\ \cite{nak12}), which is closely related to the remnant age.  Reduction in the GeV  -- TeV $\gamma$-ray luminosity in middle-aged SNRs, if made primarily by protons and ions, seems to { follow} the behavior of the { X-ray synchrotron luminosity radiated} by high-energy electrons at earlier times. { The evidence for the trend is weak, however, given that IC 443, an intermediate-aged SNR, is less luminous than most of the older remnants, in particular, W44 and W41.} The decline in the $\gamma$-ray luminosity { with age in intermediate and middle-aged SNRs, if confirmed, would instead have to be due to escape rather than energy loss as in the case of X-ray emitting electrons}, given the long time scale for energy loss of cosmic-ray protons  through inelastic nuclear processes.   Finally, we note that the assumption of a uniform medium surrounding the SNR is an extreme simplification of the actual environment of SNRs, particularly the core-collapse type found in star-forming regions.

\section{Conclusions}

Data from the new generation of $\gamma$-ray telescopes can be used to test the hypothesis that SNRs are the sources of the cosmic rays. We have constructed a simple model for $\gamma$-ray production in SNRs and find that, for parameters that are expected to apply to Type Ia and II SNe, $\approx 10$\% conversion efficiencies of directed kinetic energy to nonthermal protons or electrons are required to account for the measured $\gamma$-ray luminosities of SNRs if the radiation mechanism is secondary nuclear production or bremsstrahlung, respectively. The { small} $\gamma$-ray luminosities in young SNRs { compared to intermediate-aged SNRs} can be explained if the target material for cosmic-ray interactions is the shocked interstellar medium. Smaller efficiencies can account for the $\gamma$ rays if the emission process is Compton scattering, but the energy has to be converted to highly relativistic electrons. If the relative acceleration efficiency for electrons and protons is given by the electron-proton ratio in the cosmic rays, of order $\sim 1$\%, then the electron bremsstrahlung $\gamma$-ray luminosity would be too weak to explain the $\gamma$-ray luminosity of SNRs. 

We also looked for trends in $L_\gamma$ and the GeV to TeV flux ratios with SNR age. Except for RX J1713.7-3946 and RX~J0852.0-4622,} SNRs are found to be most $\gamma$-ray luminous at ages of $\approx 10^3$ -- $10^4$ yr. The two anomalous SNRs, RX J1713.7-3946 and RX~J0852.0-4622, have the hardest GeV-TeV spectra (Fig.\ 3) and are found in low-density environments, suggesting a Compton origin for their $\gamma$-ray emission.  The { apparent} decline of $L_\gamma$ at later times might be due to escape of cosmic rays from the acceleration and target interaction sites, but {  evidence for this trend is weak}, and the possible presence of $\gamma$-ray emission due to pulsar winds must also be carefully considered.   The tendency of the GeV to TeV flux ratio of SNRs to increase with age, or age measured in units of the characteristic Sedov age of the remnant, is suggested by this work, but will require additional analyses and study to establish.

\begin{acknowledgements}
The work of CDD is supported by the Office of Naval Research and the Fermi Guest Investigator program. The work of GP was performed through the Science and Engineering Apprenticeship Program (SEAP) at NRL. We would like to thank T.~J.\ Brandt, D.\ Horan, J.~D.\ Finke, and R.\ Yamazaki for comments on this work, and the referee for an illuminating report, suggestions, and corrections.

\end{acknowledgements}

\end{document}